\documentclass{article}

\usepackage{arxiv}

\usepackage[utf8]{inputenc} % allow utf-8 input
\usepackage[T1]{fontenc}    % use 8-bit T1 fonts
\usepackage{hyperref}       % hyperlinks
\usepackage{url}            % simple URL typesetting
\usepackage{booktabs}       % professional-quality tables
\usepackage{amsfonts}       % blackboard math symbols
\usepackage{nicefrac}       % compact symbols for 1/2, etc.
\usepackage{microtype}      % microtypography
\usepackage{lipsum}
\usepackage{amsmath}
\usepackage{cleveref}
\usepackage{graphicx}
\usepackage{adjustbox}
\usepackage{multirow}
\usepackage{multicol}
\usepackage{xcolor}
\usepackage{subfig}
\usepackage[section]{placeins}
\usepackage{}
\usepackage{pifont}% http://ctan.org/pkg/pifont

\graphicspath{ {./images/} }

\title{Meningioma segmentation in T1-weighted MRI leveraging global context and attention mechanisms}

\author{
 David Bouget \\
  Department of Medical Technology\\
  SINTEF\\
  Trondheim, Norway \\
  \texttt{david.bouget@sintef.no} \\
   \And
 Andr{\'e} Pedersen \\
  Department of Medical Technology\\
  SINTEF\\
  Trondheim, Norway \\
  \texttt{andre.pedersen@sintef.no} \\
    \And
 Sayied Abdol Mohieb Hosainey \\
  Department of Neurosurgery\\
  Bristol Royal Hospital for Children\\
  Bristol, United Kingdom \\
  \texttt{s.a.m.h@live.no} \\
    \And
 Ole Solheim \\
  Department of Neurosurgery\\
  St. Olavs hospital\\
  Trondheim, Norway \\
  \texttt{ole.solheim@ntnu.no} \\
     \And
 Ingerid Reinertsen \\
  Department of Medical Technology\\
  SINTEF\\
  Trondheim, Norway \\
  \texttt{ingerid.reinertsen@sintef.no} \\
}
\begin{document}
\maketitle
\begin{abstract}
\textbf{Purpose:} Meningiomas are the most common type of primary brain tumor, accounting for approximately 30\% of all brain tumors. A substantial number of these tumors are never surgically removed but rather monitored over time. Automatic and precise meningioma segmentation is therefore beneficial to enable reliable growth estimation and patient-specific treatment planning.\\
\textbf{Methods:} In this study, we propose the inclusion of attention mechanisms on top of a U-Net architecture used as backbone: (i) Attention-gated U-Net (AGUNet) and (ii) Dual Attention U-Net (DAUNet), using a 3D MRI volume as input. Attention has the potential to leverage the global context and identify features' relationships across the entire volume. To limit spatial resolution degradation and loss of detail inherent to encoder-decoder architectures, we studied the impact of multi-scale input and deep supervision components. The proposed architectures are trainable end-to-end and each concept can be seamlessly disabled for ablation studies.\\
\textbf{Results:} The validation studies were performed using a 5-fold cross validation over 600 T1-weighted MRI volumes from St. Olavs University Hospital, Trondheim, Norway. Models were evaluated based on segmentation, detection, and speed performances and results are reported patient-wise after averaging across all folds.
For the best performing architecture, an average Dice score of 81.6\% was reached for an F1-score of 95.6\%. With an almost perfect precision of 98\%, meningiomas smaller than 3\,ml were occasionally missed hence reaching an overall recall of 93\%.\\
\textbf{Conclusion:} Leveraging global context from a 3D MRI volume provided the best performances, even if the native volume resolution could not be processed directly due to current GPU memory limitations. 
Overall, near-perfect detection was achieved for meningiomas larger than 3\,ml which is relevant for clinical use. In the future, the use of multi-scale designs and refinement networks should be further investigated to improve the performance. A larger number of cases with meningiomas below 3\,ml might also be needed to improve the performance for the smallest tumors.  
\end{abstract}

% keywords can be removed

\section{Introduction}
\label{intro}
Primary brain tumors, characterised by an uncontrolled growth and division of cells, can be grouped into two main categories: gliomas and meningiomas. Gliomas are the brain tumors with the highest mortality rate~\cite{bauer2013survey}, while meningiomas account for one-third of all operated central nervous system tumors~\cite{ostrom2019cbtrus}. Many meningiomas are nevertheless never surgically removed, but may be encountered as incidental findings on neuroimaging. The prevalence rate of meningiomas in the general population undergoing 1.5T non-enhanced MRI scans is 0.9\%~\cite{vernooij2007incidental}, while the increase in incidence is presumably due to higher detection rates from a widespread use of Magnetic Resonance Imaging (MRI) in the general population~\cite{solheim2014effects}.
According to the EANO consensus guidelines~\cite{goldbrunner2016eano}, asymptomatic patients can be managed through observation. MRI follow-ups of benign meningiomas (i.e., WHO grade I) should be done annually, then every two years after five years. Surgery is then usually indicated if a follow-up shows tumor growth. Today, growth assessment in a clinical setting is routinely based on eyeballing or crude measures of tumor diameters~\cite{berntsen2020volumetric}. Manual segmentation by radiologists is time-consuming, tedious, and subject to intra/inter-rater variations difficult to characterize~\cite{binaghi2016collection} and is therefore rarely done in clinical routine. Systematic and consistent brain tumor segmentation and measurements through (semi-)automatic methods are therefore of utmost importance. Patient-specific follow-up plans could potentially be enabled by an increased sensibility of growth measures and estimation of future growth.
In addition, assessing the growth pattern on individual level may be informative with respect to treatment indication, as a majority may exhibit a self-limiting growth pattern~\cite{behbahani2019prospective}. In addition, segmentation is key for assessing treatment responses after radiotherapy or surgery.
MRI represents the gold standard for medical imaging due to its non-invasiveness and widespread availability, most often using contrast-enhanced T1-weighted sequences where the tumor border is more easily distinguishable~\cite{goldbrunner2016eano}. Alternatively, the fluid-attenuated inversion recovery (FLAIR) sequence can complement the diagnosis by better capturing the edema region for the cerebrospinal fluid. Nevertheless, inherent downsides can be associated with MRI acquisitions such as intensity inhomogeneity~\cite{tustison2010n4itk}, variations from the use of different acquisition scanners~\cite{nyul2000new}, or variations in the acquisitions (e.g., field-of-view, slice thickness, or resolution). In T1-weighted MRI, meningiomas are often sharply circumscribed with a strong contrast enhancement, making them clear to identify. However, small meningiomas might resemble other contrast-enhanced structures such as blood vessels, hindering the detection task. In order to alleviate radiologists' burden to annotate large contrast-enhanced meningiomas, while at the same time to help detecting smaller and unusual meningiomas, automatic segmentation methods are paramount.

\setlength{\parindent}{5ex} In recent years, automatic and end-to-end semantic segmentation has known considerable improvements through the development of encoder-decoder fully convolutional neural network architectures (FCNs)~\cite{badrinarayanan2017segnet,long2015fully,ronneberger2015u}. By restoring the feature map of the last deconvolution layer to the size of the initial input sample, predictions can be generated for each voxel. While such architectures provide near radiologist-level performances on some medical image analysis tasks~\cite{bai2017human,liao2019evaluate}, multi-stage cascading induces a loss of local information leading to excessive and redundant low-level features. The most effective solution to boost the segmentation performance is to combine local and global information to preserve consistency in the feature maps.
However, 3D medical volumes are typically too sizable to fit on GPU memory at their original resolution, and the amount of parameters for the corresponding model would be considerable. Different trade-offs have been investigated such as splitting the 3D volume into a series of patches or slabs by which some global context can be leveraged while good local information is retained~\cite{akkus2017deep}. Capturing the entire global context from a full 3D volume is important for a model to understand the spatial relationships between the different anatomical structures.
Aggregating multi-scale contexts and using various dilated convolutions and pooling operations~\cite{chen2017rethinking,zhao2017pyramid}, enlarging kernel size to capture richer global information~\cite{peng2017large}, or fusing semantic features at different levels~\cite{lin2017refinenet} can cope with information loss but are unable to leverage overall relationships between structures. To address shortcomings from feature maps consistency and loss of information when using multi-stage cascading architectures, attention mechanisms have been utilized with great success~\cite{fu2019dual,oktay2018attention,cheng2020fully}. Attention modules can be seamlessly coupled with regular FCN architectures for end-to-end training, with the benefit of letting the model learn to focus on the segmentation targets without significant computational overhead nor additional model parameters. Optimally coupled with each deconvolution block, attention can be designed to capture features' dependencies spatially, channel-wise, or across any other dimension~\cite{fu2019dual}.
Alternatively, multiple models operating on different input shapes or focusing on different aspects during training can be fused as a post-processing step, called an ensemble, to generate the final prediction map~\cite{feng2020brain}. Global context and local refinement can virtually be obtained separately at the cost of longer training and inference time, and higher model complexity. However, ensembling has not always shown to produce better overall segmentation performance compared to a single model's use~\cite{aresta2019bach}.

\setlength{\parindent}{5ex} Few studies have investigated the task of meningioma segmentation in-depth, more often focusing on gliomas given the open access to the BRATS challenge dataset~\cite{menze2014multimodal}. A combination of support vector machine and graph cut has been used for multi-modal and multi-class segmentation, but results were reported on a meager set of 15 patients~\cite{binaghi2018meningioma}. In consecutive studies, Laukamp et al. used different neural network architectures (e.g., DeepMedic, BioMedIA) over 3D volumes on their own multi-modal dataset~\cite{laukamp2019fully,laukamp2020automated}. Good segmentation performance was obtained, but heavy preprocessing techniques such as atlas registration and skull-stripping were performed. In their study, Pereira et al. reported benefits from normalization and simple data augmentation for brain tumor segmentation~\cite{pereira2016brain}. In our previous work, leveraging a whole MRI volume, rather than slab-wise, has shown to boost the overall segmentation performance~\cite{bouget2020fast}. However, using a regular 3D U-Net or multi-scale architecture still resulted in the loss of information in the encoding path which remained to be addressed.
Over the last years, almost all brain tumor segmentation studies were based around deep learning~\cite{icsin2016review}. Simple 3D CNN architectures~\cite{myronenko20183d,isensee2018no}, multi-scale approaches~\cite{kamnitsas2017efficient,xu2018multi}, and ensemble of multiple models~\cite{feng2020brain} have been explored. While they achieve better segmentation performance, are more robust to hyperparameters, and generalize better; the potential from using a whole 3D MRI volume remains yet to be fully explored.

\setlength{\parindent}{5ex} In this paper, we focus on reducing the information loss for encoder-decoder architectures using combinations of attention, multi-scale, and deep supervision schemes. Our contributions are: (i) the investigation of architectures able to better understand global context, (ii) validation studies focusing on meningioma volumes for clinical and diagnostic use, and (iii) online availability for our trained models along with the inference script.

\section{Methods}
\label{sec:methods}

\subsection{Related work}
Typically, semantic features are extracted along the encoding path for encoder-decoder architectures. The field-of-view is progressively enlarged via strided convolutions or pooling operations, hence provoking some loss of detail. In the decoding path, extracted features are exploited in order to solve the task at hand (i.e., classification, segmentation). At the end of the encoding path, the feature maps are the richest in global relationships, but limited spatial details are preserved due to cascaded convolutions and nonlinearities. In order to recover fine-grained details, symmetrical architectures, such as U-Net, propagate feature maps across corresponding encoder and decoder at the same level, also known as long skip connections. In general, efficient architectures optimally use global and contextual information from high-level features and border information from low-level features to resolve small details~\cite{sang2020pcanet}.
Attention mechanisms focus on identifying salient image regions and amplifying their influence while filtering away irrelevant and potentially confusing information from other regions, making the prediction more contextualised~\cite{jetley2018learn}. Hard attention, stochastic and non-differentiable, relies on sampling-based training making optimizing models more difficult. Soft attention, probabilistic and amenable to training by backpropagation, can be by contrast seamlessly integrated into current CNN architectures, which has been applied in numerous tasks such as text understanding and semantic segmentation~\cite{lin2017structured,vaswani2017attention,lin2016efficient}.

\setlength{\parindent}{5ex} In a main body of work, a single attention gating is performed at every level along the decoding path. Attention feature maps are often concatenated with the feature maps from the long skip connection~\cite{oktay2018attention, abraham2019novel}, but propagating the lowest-level feature maps in an upward fashion with short skip connection has also been investigated for computing the attention feature maps~\cite{schlemper2019attention}. In a second body of work, authors have investigated the computation of specific attention feature maps to focus on position, channel, or class dependencies. Fu et al. presented a dual attention network for scene segmentation where position and channel attention modules are computed at the bottom of a ResNet encoding path~\cite{fu2019dual}. The generation of the final probability map, right after and without a matching decoding path, is detrimental to the spatial segmentation quality. Following the same idea, Mou et al. added a complete ResNet decoding path after position and channel attention computation, improving the spatial reconstruction~\cite{mou2019cs}. Attempts have been made to include dual attention modules at every stage of a ResNet architecture, either from the skip connection feature maps from the encoder path~\cite{cheng2020fully}, or in the decoder path after concatenation with the feature maps from the previous level~\cite{sinha2020multi}. The use of dilated convolutions, or the addition of a significant dropout over the attention feature maps, appear necessary to deal with the substantial amount of parameters and prevent training hurdles (e.g., overfitting, slow convergence). Finally, other hybrid attention schemes have been explored, for example in the context of aerial image segmentation with concepts such as class channel attention to exploit dependencies between classes and generate class affinity maps~\cite{niu2020hmanet}.

\setlength{\parindent}{5ex} To compensate for the loss of detail inherent to consecutive pooling operations, new architecture designs or layers have been proposed. In order to preserve details in the encoding path, various multi-scale attempts have been made, such as infusing down-sampled version of the input volume in each encoder block~\cite{abraham2019novel}, using atrous convolutions and pyramid spatial pooling to enlarge the receptive fields~\cite{chen2017rethinking,sang2020pcanet}, or by concatenating the feature maps from each encoder block and using the created multi-scale feature maps for guiding in an upward skip connection fashion~\cite{sinha2020multi}. In the latter case, complementary low-level information and high-level semantics are encoded jointly in a more powerful representation.
Conversely, intermediate feature maps generated at each level of an encoder-decoder architecture can be leveraged instead of computing the loss simply from the last decoder step, commonly referred to as Deep Supervision (DS). The rationale is that the feature maps from hidden layers of a deep network can serve as a proxy to improve the overall segmentation quality and sensitivity of the model, while alleviating the problem of vanishing gradients~\cite{lee2015deeply}. The final loss is computed as a weighted average between the losses from each level whereby each can contribute equally~\cite{abraham2019novel}, or with weights defined as trainable parameters. Intermediate losses can be computed separately from the raw feature maps and the attention feature maps, before tallying the final loss across all levels~\cite{sinha2020multi}. In general, the combination of multi-resolution and deep supervision has shown to improve convergence (i.e., better optimum and faster solving) for inverse problems~\cite{wang2020multi}.

\subsection{Dataset}
In a previous study~\cite{bouget2020fast}, we introduced a dataset of 698 Gd-enhanced T1-weighted MRI volumes acquired on 1.5T and 3T scanners in the catchment region of the Department of Neurosurgery at St. Olavs University hospital, Trondheim, Norway. In this study, we kept the 600 high-resolution MRI volumes having a maximum spacing of 2\,mm along the z-axis. Of those 600 patients, 276 underwent surgery to resect the meningioma, while the remaining 324 were followed at the outpatient clinic. In the dataset, MRI volume dimensions covered $[240; 512]\times[224; 512]\times[18; 290]$ voxels and the voxel sizes ranged between $[0.47; 1.05]\times[0.47; 1.05]\times[0.60; 2.00]\,\text{mm}^3$. 
The volumes of the surgically resected meningiomas were on average larger ($30.92\pm33.10\,\text{ml}$), compared to the untreated meningiomas followed at the outpatient clinic ($7.62\pm13.67\,\text{ml}$). Overall, meningioma volumes ranged between $[0.07, 167.99]\,\text{ml}$ for an average value of $18.33\pm27.20\,\text{ml}$.

\begin{figure}[!ht]
\centering
\includegraphics[scale=0.5]{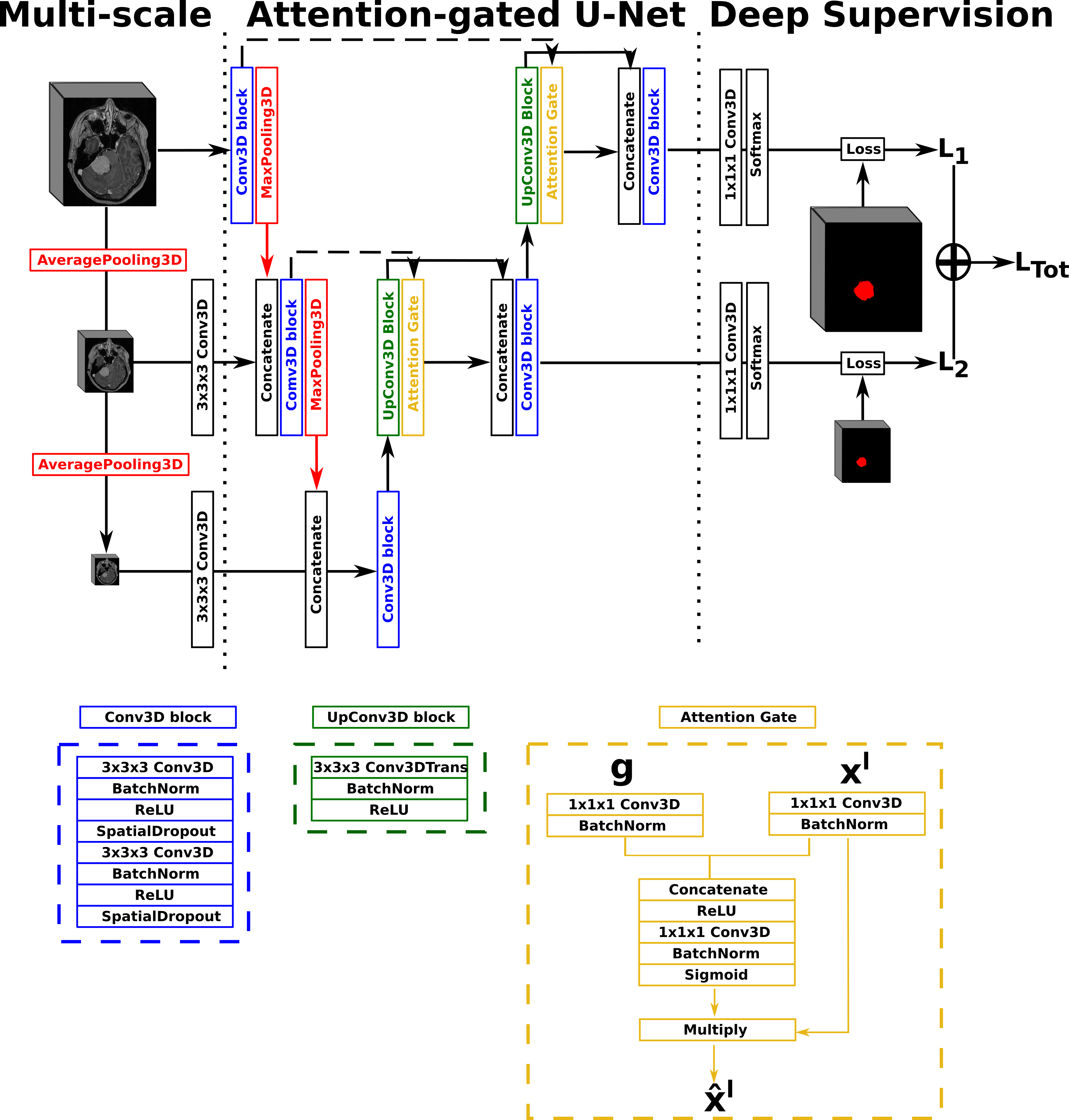}
\caption{Illustration of the Attention-Gated U-Net (AGUNet), with multi-scale input and deep supervision. The representation features three levels for viewing purposes, but five levels have been used in the studies.}
\label{fig:unet-arch}
\end{figure}

\subsection{Architecture design}
\label{subsec:preproc}
In this work, we opted for a U-Net architecture as backbone, which we set to five levels and used filter sizes of $[16, 32, 128, 256, 256]$ for each level, respectively. Our first proposed architecture, named AGUNet and illustrated in Fig.~\ref{fig:unet-arch}, integrates an attention-gated mechanism to U-Net. Our second architecture, named DAUNet and illustrated in Fig.~\ref{fig:dualunet-arch}, integrates a dual attention module to U-Net. In addition, both architectures are combining multi-scale input and deep supervision support.
For viewing purposes and clarity, we chose to display our proposed architectures with only three levels. The proposed design is modular whereby the backbone can be changed (e.g., U-Net, ResNet) and each main module (i.e., multi-scale input, attention mechanism, and deep supervision) can be toggled, hence enabling seamless end-to-end training while providing unbiased and comparable results. The specifics of each module are presented in the following subsections.

\subsubsection{Attention mechanisms}
In our first architecture, attention gates were incorporated to each decoder step of the architecture in order to highlight salient features passing through the skip connections, as described previously~\cite{oktay2018attention}. The attention gating is performed before the concatenation operation in order to merge only relevant activations, disambiguated from irrelevant responses in skip connections by performing gating from information extracted at a coarser scale. At each decoder level, the feature maps from the previous level (i.e., coarser scale) are first resampled to match the shape of the skip connection feature maps, using a transpose convolution operation with a $3\times3\times3$ kernel size (cf. green block in Fig.~\ref{fig:unet-arch}). Inside the attention gate (cf. yellow block in Fig.~\ref{fig:unet-arch}), the upsampled feature maps (denoted as $g$) and the feature maps from the skip connection (denoted as $x^l$) are processed to generate the gating signal which is applied to $x^l$ in order to generate the gated feature maps for the current level $\hat x^l$. Linear transformations without spatial support (i.e., $1\times1\times1$ convolutions) are performed to limit the computational complexity and amount of trainable parameters, similarly to non-local blocks~\cite{wang2018non}. We chose to include an attention gate on the lowest-level feature maps (i.e. first skip connection) even though limited benefits are expected since the input data tend to not be represented in a high-enough dimensional space~\cite{jetley2018learn}.

\begin{figure}[ht]
\centering
\includegraphics[scale=0.5]{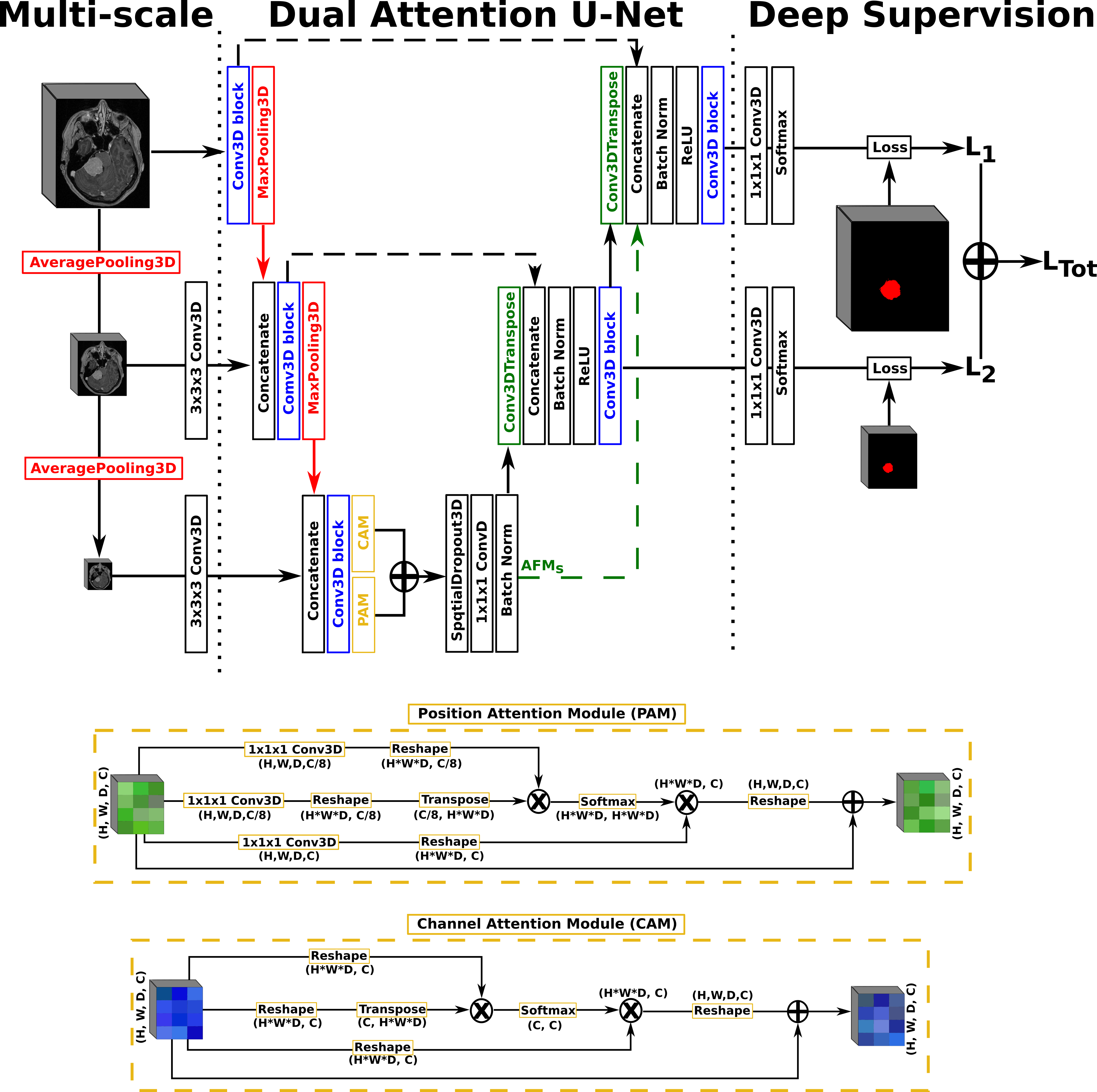}
\caption{Illustration of the Dual Attention U-Net (DAUNet), with multi-scale input, deep supervision, and the same Conv3D block as described in the first architecture. The representation features three levels for viewing purposes, but five levels have been used in the studies.}
\label{fig:dualunet-arch}
\end{figure}

\setlength{\parindent}{5ex} As second architecture, a dual attention scheme with position and channel attention modules was integrated to the U-Net architecture. Due to GPU-memory limitations and to reduce the computational complexity, the attention feature maps are only computed once at the end of the encoding path rather than at every decoder level. The position attention module, or spatial attention module, encodes a wider range of contextual information into local features to enhance their representation capability. The channel attention module exploits inter-dependencies between channel maps to emphasize inter-dependant feature maps and improve the feature representation of specific semantics, as presented by Fu et al.~\cite{fu2019dual}. For training efficiency, a spatial dropout operation with a rate of 0.5 and linear transformations are performed on the raw attention feature maps, generating the final Attention Feature Maps (AFMs).
In a variant, named DAGUNet, the attention feature maps are propagated upward and concatenated at each decoder level to guide the computation of feature maps at the higher levels (cf. green arrow in Fig.~\ref{fig:dualunet-arch}). Transferring the bottom attention feature maps requires less trainable parameters overall than computing the dual attention blocks at every decoder step, while still benefiting from them.

\subsubsection{Multi-scale and deep supervision}
For our multi-scale approach, we opted to exploit down-sampled versions of the initial network input, at every level in the encoding path, by performing consecutive average pooling operations with a $3\times3\times3$ kernel size. Each down-sampled volume is then concatenated to the feature maps coming from the previous encoding level, before generating the feature maps for the current level, in order to preserve spatial details.
For our deep supervision scheme, the ground truth volume is recursively down-sampled to match the size of the feature maps at each corresponding decoder level, where an intermediate loss $L_{x}$ is computed. The final loss, represented as $L_{\text{Tot}}$ in Fig.~\ref{fig:unet-arch}, is the weighted sum from all intermediate losses. In this study, we did not set the weights as trainable parameters, not to favor the feature maps from any level, and kept a uniform weighting strategy.

\subsection{Training strategies}
\label{subsec:train-strats}
The MRI volumes were all preprocessed using the following steps: (i) resampling to an isotropic spacing of $1\,\text{mm}^3$ using spline interpolation order 1 from NiBabel~\footnote{https://github.com/nipy/nibabel}, (ii) clipping tightly around the patient's head, (iii) volume resizing to $128\times128\times144\,\text{voxels}$ using spline interpolation order 1, and (iv) normalizing intensities to the range $[0, 1]$.
A typical data augmentation approach was favored, where the following transforms were applied to each input sample with a probability of 50\%: horizontal and vertical flipping, random rotation in the range $[-20^{\circ}, 20^{\circ}]$, translation up to 10\% of the axis dimension, zoom between $[80, 120]\%$ in the axial plane.

\setlength{\parindent}{5ex} All models were trained from scratch using the Adam optimizer with an initial learning rate of $10^{-3}$ and training was stopped after 30 consecutive epochs without validation loss improvement. The main loss function used was the class-average Dice loss, excluding the background. Additionally, we experimented with the Focal Tversky Loss (FTL), where the Tversky similarity index helps balance false positive and false negative predictions more flexibly, while the focal aspect increases the contribution of hard training examples in the loss computation~\cite{abraham2019novel}. We used $\alpha=0.7$ and $\beta=0.3$ for the Tversky index to minimize false negative predictions, and $\gamma=2.0$ as focal parameter. Unless specified otherwise, all models were saved based on the minimum overall validation loss, which corresponds to $L_{\text{Tot}}$ if deep supervision is enabled.

\setlength{\parindent}{5ex} Given the sizable memory footprint, all models were trained using two samples in a batch. In order to improve generalization, and because mini-batches up to 32 elements have shown to produce better models~\cite{masters2018revisiting}, we used the concept of accumulated gradients to effectively increase the batch size. For a specified number of accumulated gradient steps (n), each batch is run sequentially using the same model weights for calculating the gradients. When the n steps are performed, the accumulated sum of gradients from each step amounts to the same gradients as if computed over the larger batch size, ensuring that the model weights are properly updated. For our studies, we chose to perform 16 steps, enabling us to use a batch size of 32.

\section{Validation studies}
\label{sec:validation}
In this work, we focus on maximizing the segmentation and detection performance while investigating runtime and potential for diagnostic purposes and clinical use. A 5-fold cross-validation approach was followed whereby at every iteration three folds were used for training, one for validation, and one for testing. Each fold was populated in order to exhibit a similar meningioma volume distribution, as described in our previous study~\cite{bouget2020fast}.

\textit{Metrics and measurements:}
For quantifying the performance, we used: (i) the Dice score, (ii) the F1-score, and (iii) the training/inference speed.
The Dice score is used to assess the quality of the pixel-wise segmentation (in \%), while the F1-score assesses the harmonic average of recall and precision (in \%). Finally, the training speed, the inference speed, and the total processing speed to generate results for a new MRI volume, are all reported in seconds.
For the segmentation task, the Dice score is computed between the ground truth and a binary representation of the probability map generated by a trained model. The binary representation is computed for ten different equally-spaced probability thresholds (PT), in the range $[0, 1]$.
A connected components approach, coupled to a pairing strategy, was employed to compute the recall and precision values. Such step is mandatory for the minority of multifocal meningiomas, but also to separate the correct prediction over a meningioma from the false positives per patient, enabling to also report the true positive Dice (Dice-TP). 
Pooled estimates, computed from each fold's results, are computed for each measurement~\cite{killeen2005alternative}, and reported with mean and standard deviation.

\textit{(i) Ablation study:} Comparison of segmentation performances using various combinations of the methodological components introduced in Section~\ref{sec:methods}, whereby the name given to each experiment is a concatenation of components' abbreviations. The architectures to choose from are: regular U-Net (UNet), attention-gated U-Net (AGUNet), dual attention U-Net (DAUNet), and dual attention guided U-Net (DAGUNet), combined with multi-scale input (MS), deep supervision (DS), and the use of accumulated gradients (AG). If not specified otherwise, the Dice loss function is used and the best model is selected based on the total loss $L_{\text{Tot}}$. Usage of the focal Tversky loss is indicated by the TFL tag, while saving the best model based on the loss from the upper level is indicated by the Top tag.

\textit{(ii) Segmentation and detection performances study:} A comparison is performed between the best trained model for each of the main designs: slab-wise U-Net (UNet-Slabs) and PLS-Net studied previously~\cite{bouget2020fast}, full volume U-Net (UNet-FV) and the best method identified in the ablation study (Ours). All models were compared using the exact same methodology considering only the probability threshold PT, without any consideration toward the absolute size or relative overlap of the meningioma candidates.

\textit{(iii) Volume-based performances analysis:} To study the potential for clinical use in the hospital or outpatient clinic, performances are analyzed over different meningioma groups based on volume. Limitations such as challenging meningiomas and potential outliers are also described.

\textit{(iv) Speed performances study:} For the different experiments considered in the first two validation studies, a deeper analysis around speed is conducted. The model complexity as total number of parameters and the training behaviour as $s.epoch^{-1}$ (in seconds), best epoch, and total training time (in hours), are first considered. The pure inference speed is reported when using GPU and CPU (in s). Finally, the total elapsed time required to generate predictions for a new patient is reported as processing time (in s), obtained with GPU support. The operations required to prepare the data to be sent through the network, to initialize the environment, to load the trained model, and to reconstruct the probability map in the referential space of the original volume are accounted for. The experiment has been repeated 10 consecutive times over the same MRI volume for each model, using a representative sample of $256\times256\times192$\,voxels with $1.0\times1.0\times1.0$\,mm spacing.

\section{Results}
\label{sec:results}
Models were trained across different machines using either an NVIDIA Quadro P5000 (16GB) or a Tesla P100 PCIe (16GB) dedicated GPU, and regular hard-drives. For inference and processing speed computation, an Intel Xeon @3.70 GHz (6 cores) CPU and an NVIDIA Quadro P5000 GPU were used. Implementation was done in Python 3.6 using \texttt{TensorFlow} v1.13.1, Cuda 10.0, and the Imgaug Python library for the data augmentation methods~\cite{imgaug}. Due to randomness during weight initialization and optimization, a fixed seed was set to make comparisons between experiments fair and reproducible. Trained models and inference code are made publically available at \url{https://github.com/dbouget/mri_brain_tumor_segmentation}.

\subsection{Ablation study}
 
\begin{table}[h]
\caption{Performances obtained by component ablation, averaged over the five folds. The components are: regular U-Net (UNet), attention-gated U-Net (AGUNet), dual attention U-Net (DAUNet), dual attention guided U-Net (DAGUNet), multi-scale input (MS), deep supervision (DS), and the use of accumulated gradients (AG).}
\adjustbox{max width=\textwidth}{
\begin{tabular}{l|c|c|c|c|c|c|c|c}
Experiment & PT & Dice & Dice-TP & F1 & Recall & Precision \tabularnewline
\hline
(i) UNet-FV & 0.5 & $76.91\pm28.98$ & $84.77\pm16.22$ & $93.19\pm01.70$ & $90.70\pm01.90$ & $95.86\pm02.28$ \tabularnewline
(ii) AGUNet-AG & 0.4 & $75.14\pm30.38$ & $84.21\pm16.57$ & $92.15\pm01.74$ & $89.21\pm02.38$ & $95.35\pm02.37$ \tabularnewline
(iii) AGUNet-DS-AG & 0.4 & $80.72\pm24.98$ & $86.79\pm12.19$ & $94.73\pm00.76$ & $93.02\pm02.01$ & $96.63\pm02.95$ \tabularnewline
(iv) AGUNet-MS-DS & 0.4 & \boldmath{$81.64\pm25.33$} & \boldmath{$87.69\pm12.12$} & \boldmath{$95.58\pm02.24$} & $93.03\pm04.13$ & \boldmath{$98.39\pm01.43$} \tabularnewline
(v) AGUNet-MS-DS-AG & 0.4 & $79.49\pm26.38$ & $87.02\pm11.59$ & $94.23\pm00.88$ & $91.69\pm01.73$ & $96.93\pm01.04$ \tabularnewline
(vi) AGUNet-MS-DS-AG-Top & 0.5 & $79.89\pm26.52$ & $86.64\pm13.75$ & $94.53\pm08.23$ & $92.19\pm02.21$ & $97.07\pm01.67$ \tabularnewline
(vii) AGUNet-MS-DS-AG-FTL & 0.7 & $74.27\pm30.29$ & $84.21\pm14.47$ & $91.27\pm01.50$ & $88.20\pm02.96$ & $94.66\pm01.94$ \tabularnewline
(viii) DAUNet-MS-DS-AG & 0.5 & $78.43\pm27.56$ & $85.92\pm13.73$ & $92.99\pm02.76$ & $91.19\pm04.71$ & $95.04\pm02.89$ \tabularnewline
(ix) DAGUNet-MS-DS & 0.4 & $81.54\pm24.95$ & $87.15\pm13.34$ & $95.24\pm01.33$ & \boldmath{$93.52\pm02.39$} & $97.06\pm00.83$ \tabularnewline
(x) DAGUNet-MS-DS-AG & 0.4 & $80.74\pm24.89$ & $86.79\pm12.00$ & $94.78\pm00.99$ & $93.03\pm01.91$ & $96.63\pm00.76$ \tabularnewline
\end{tabular}
}
\label{tab:results-ablation-study}
\end{table}

Patient-wise segmentation and detection performances for the different architectural designs considered are summarized in Table~\ref{tab:results-ablation-study}, where the first row provides baseline results using the backbone architecture only. The greatest impact comes from the deep supervision component increasing the Dice score by about 5\% and the F1-score by 2.5\% between experiments (ii) and (iii). From the multi-scale input approach, less than 1\% improvement for the same metric is reported, as can be seen between experiments (iii) and (iv). It is worth mentioning that models trained using deep supervision produce comparable results whether saved based on the best total loss or the best loss from the upper level only (cf. experiments (v) and (vi)). The use of attention modules does not further improve the results (cf. experiments (i) and (ii)). Similarly, no added value has been recorded when using a more complex dual attention scheme (experiment x), rather than attention gating (experiment v). A similar conclusion can be drawn for the use of the accumulated gradients strategy, degrading slightly the overall segmentation performances, for a reduction in standard deviation across detection results (cf. experiments (iv) and (v)). While the implementation seems correct, identifying the best batch size is difficult and heavily dependant on the dataset size and diversity. However, for our second architecture with dual attention, propagating the attention feature maps upward seems to be beneficial with a 1-2\% increase across the different measurements when compared to no propagation (cf. experiments (viii) and (x)). The attempt to use the Focal Tversky loss instead of Dice loss was not conclusive as all metrics are lower in experiment (vii) compared to experiment (v).

\setlength{\parindent}{5ex} Overall, we consider the best performing model to be obtained by experiment (iv), reaching the highest scores for all but one metric. As we do favor detection performances over pixel-wise segmentation accuracy, our AGUNet-MS-DS model is also reaching the highest F1-score with 95.58\%. In the rest of the paper, we refer to AGUNet-MS-DS (experiment (iv)) as Ours.
 
\subsection{Segmentation and detection performances study}
\begin{table}[!b]
\caption{Segmentation and detection performances obtained with the four main designs considered, averaged over the five folds. The first two designs were introduced and detailed in our previous study~\cite{bouget2020fast}.}
\adjustbox{max width=\textwidth}{
\begin{tabular}{l|c|c|c|c|c|c|c|c|c}
Experiment & PT & Dice & Dice-TP & F1 & Recall & Precision \tabularnewline
\hline
UNet-Slabs & 0.6 & $74.41\pm29.04$ & $81.72\pm18.19$ & $82.74\pm02.65$ & $91.04\pm03.87$ & $75.91\pm02.89$ \tabularnewline
PLS-Net & 0.5 & $71.69\pm33.41$ & $83.46\pm17.96$ & $89.87\pm01.79$ & $85.88\pm03.02$ & $94.31\pm01.03$ \tabularnewline
UNet-FV & 0.5 & $76.91\pm28.98$ & $84.77\pm16.22$ & $93.19\pm01.70$ & $90.70\pm01.90$ & $95.86\pm02.28$ \tabularnewline
Ours & 0.4 & \boldmath{$81.64\pm25.33$} & \boldmath{$87.69\pm12.12$} & \boldmath{$95.58\pm02.24$} & \boldmath{$93.03\pm04.13$} & \boldmath{$98.39\pm01.43$} \tabularnewline
\end{tabular}
}
\label{tab:results-segperf-overall}
\end{table}

For the four different training concepts considered, segmentation performances have been reported in Table~\ref{tab:results-segperf-overall}. The UNet-Slabs approach yields surprisingly competitive recall performances with only a 2\% shortfall compared to our best-performing method. However, an inherent limitation of slabbing a 3D volume is the probable generation of a larger amount of false positives per patient, as can be seen by the 20\% difference in precision between the same two approaches. While the PLS-Net architecture drastically increases the precision from leveraging a full 3D volume, its shallow architecture is not able to compete in terms of overall pixel-wise segmentation or recall performances. Nevertheless, it indicates how well global spatial relationships can be modelled and how beneficial it can be for a 3D segmentation task. The simple U-Net architecture over an entire 3D volume (UNet-FV), building upon the strengths of UNet-Slabs and PLS-Net, boosts performances in every aspect. Employing advanced mechanisms such as attention, deep supervision, or multi-scale input provides slight improvements in detection performances, going from an F1-score of 93.2\% up to 95.6\%. Yet, the highest benefit can be witnessed for the pixel-wise segmentation task, with an overall Dice score reaching 81.64\%, up by almost 5\%.

\begin{figure}[!t]
\centering
\includegraphics[scale=0.85]{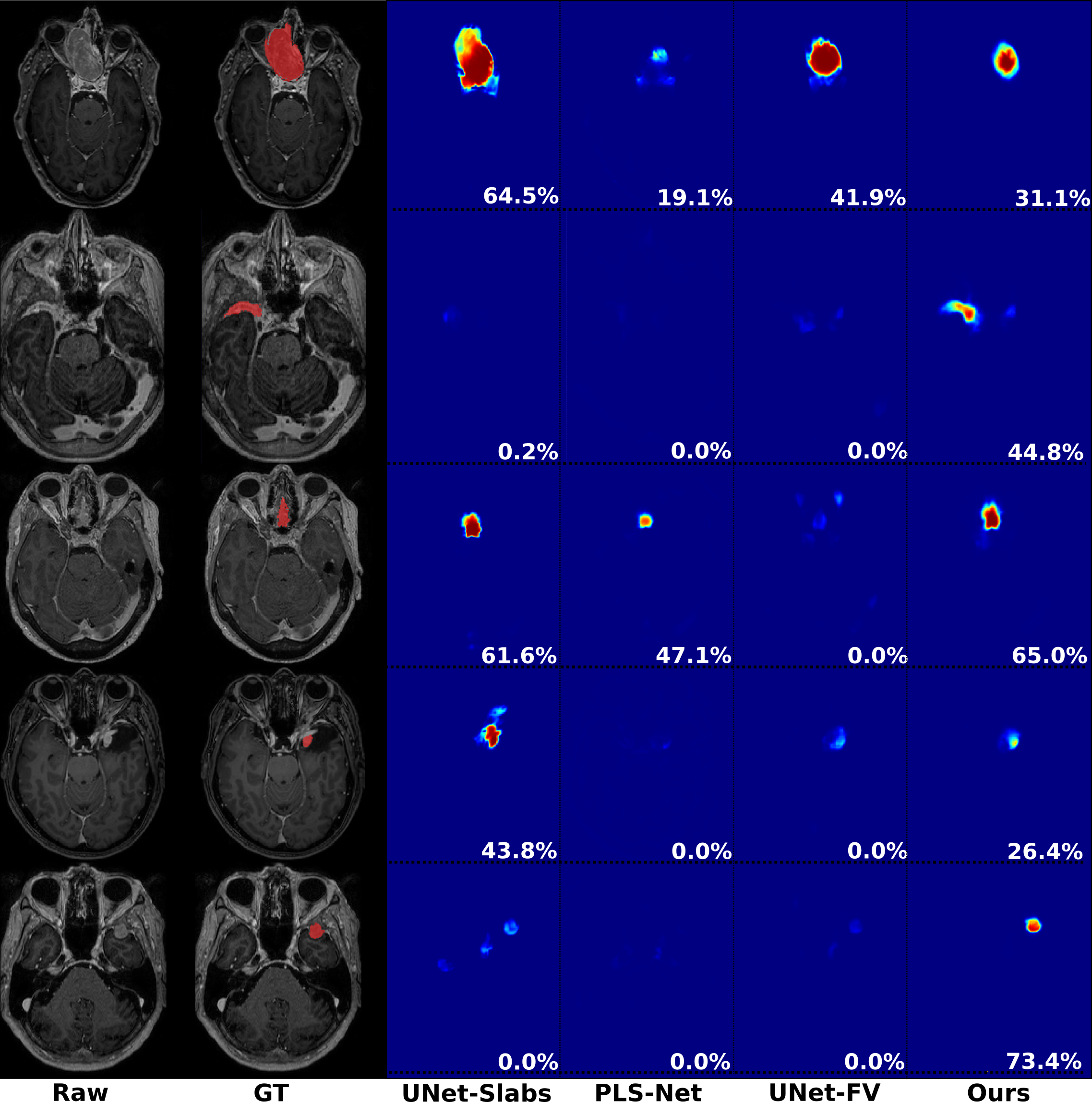}
\caption{Prediction examples for a patient from each fold, one per line. The raw Dice is reported, in white, for the optimal probability threshold identified for each model.}
\label{fig:seg-diff-overall-illu}
\end{figure}

Visual comparisons are provided in Fig.~\ref{fig:seg-diff-overall-illu} between the four methods for five different patients, one per row, featuring meningiomas located in uncommon regions of the brain or with a small volume. For the first patient displayed, the meningioma is almost completely outside the brain and has grown between the eye sockets up the back of the nose, location relatively rare and under-represented in our dataset. As such, the best Dice score is obtained with the UNet-Slabs approach, which brings more focus to local intensity gradients. The use of more global information with the UNet-FV approach reduces prediction probabilities further away from the brain, while the use of attention mechanisms with our best approach lowers the Dice score to half of the UNet-Slabs value. Even though the impact of attention mechanisms could not be overall witnessed from the values reported in Table~\ref{tab:results-ablation-study}, the results on this patient represent a perfect exemplification. Attention mechanisms seem to have learned to limit the predictions within the brain or its outskirts from the training examples, as not many meningiomas in our dataset are outgrowing this far from the brain. The patients featured in the second and fifth rows are representative for difficult meningiomas that are either small or oddly located, where only our best approach is able to perform a reasonable pixel-wise segmentation.

\subsection{Volume-based performances analysis}

\begin{figure}[!b]
\centering
\subfloat[][UNet-Slabs]{\includegraphics[scale=0.48]{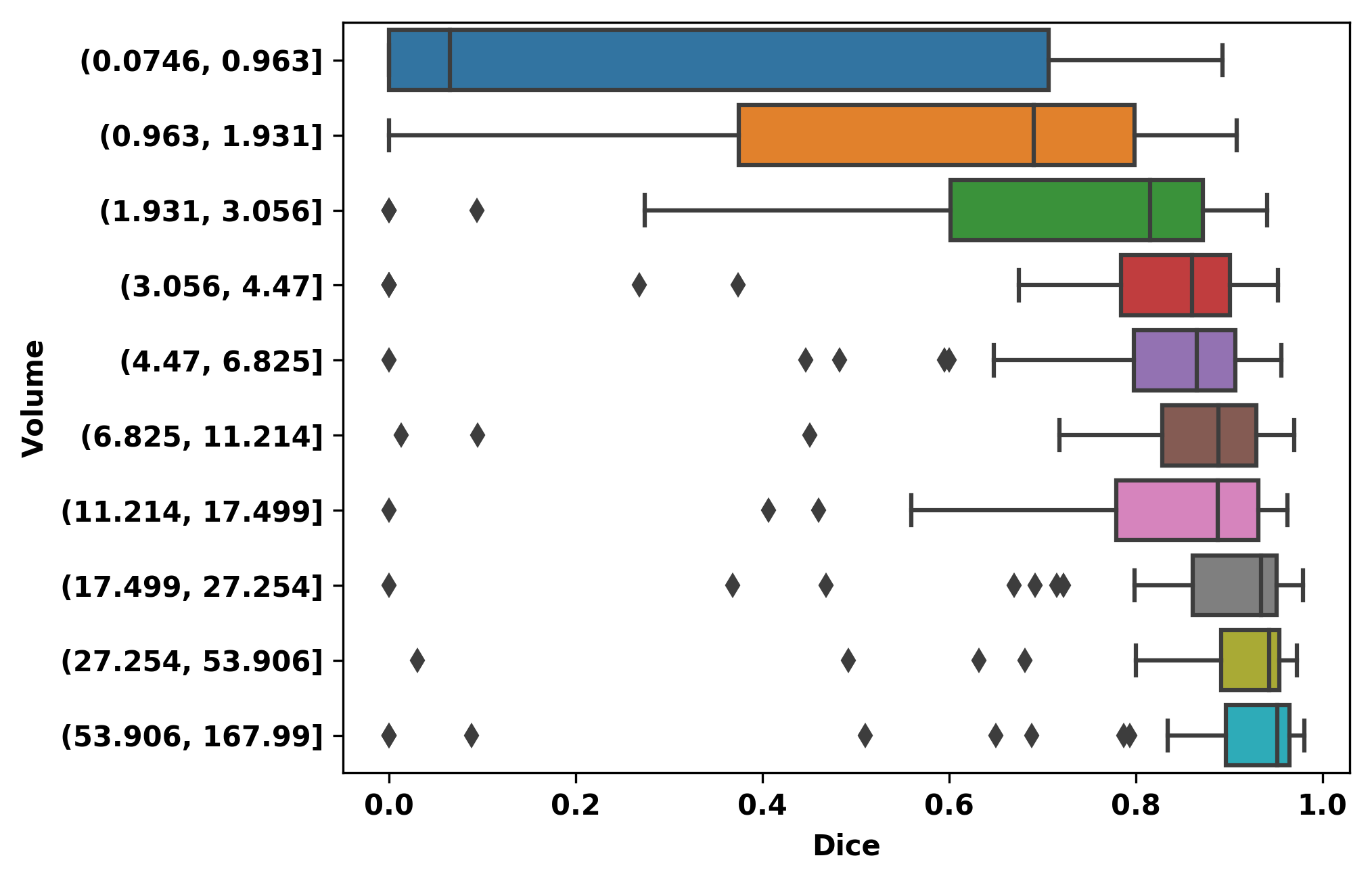}}~
\subfloat[][PLS-Net]{\includegraphics[scale=0.48]{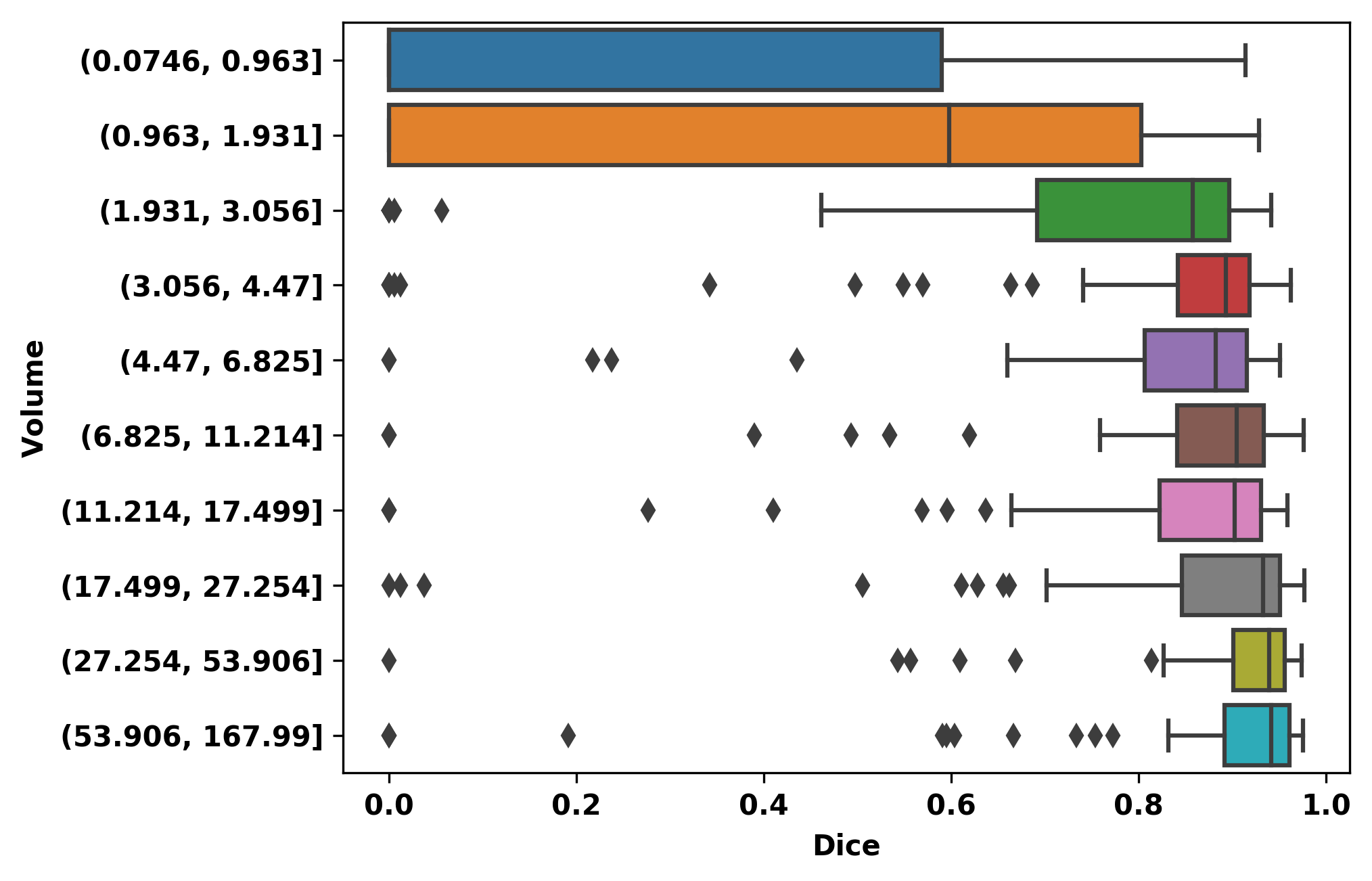}}\\
\subfloat[][UNet-FV]{\includegraphics[scale=0.48]{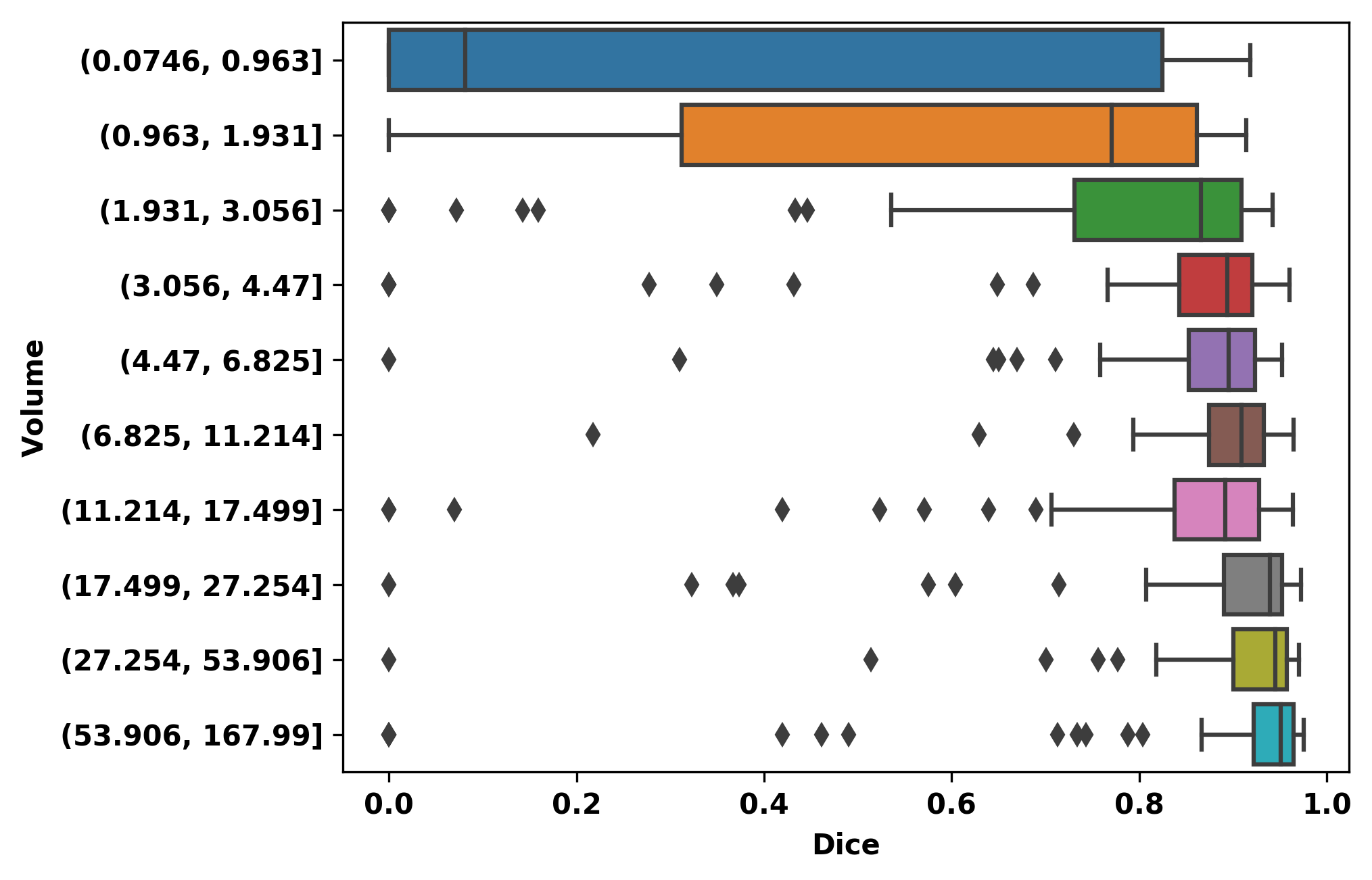}}~
\subfloat[][Ours]{\includegraphics[scale=0.48]{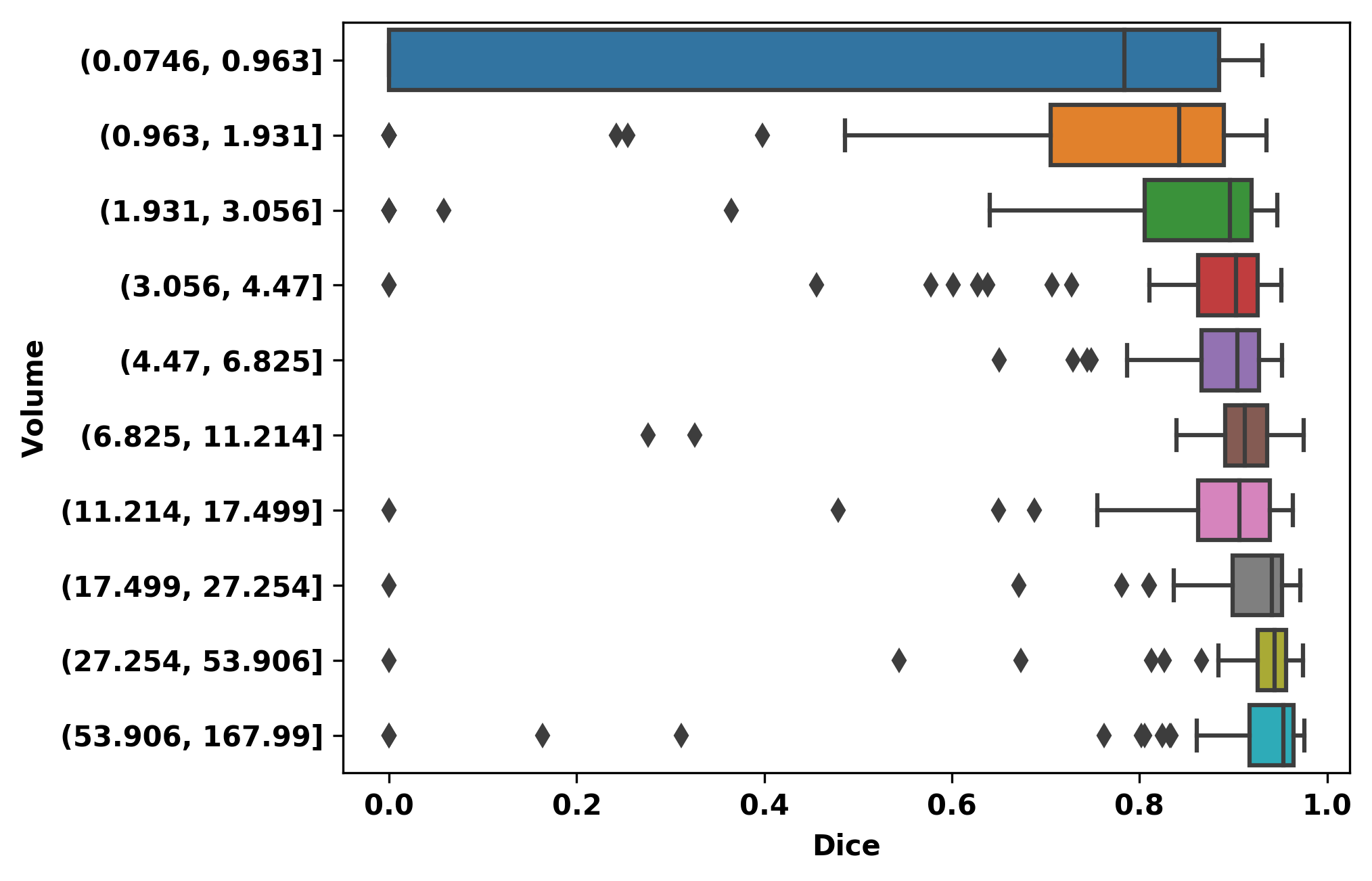}}
\caption{Segmentation performances for the four main designs represented with box plots. Ten equally-populated bins, based on tumor volumes, have been used to group the meningiomas.}
\label{fig:results-volumeprov}
\end{figure}

\begin{figure}[!t]
\centering
\includegraphics[scale=1.34]{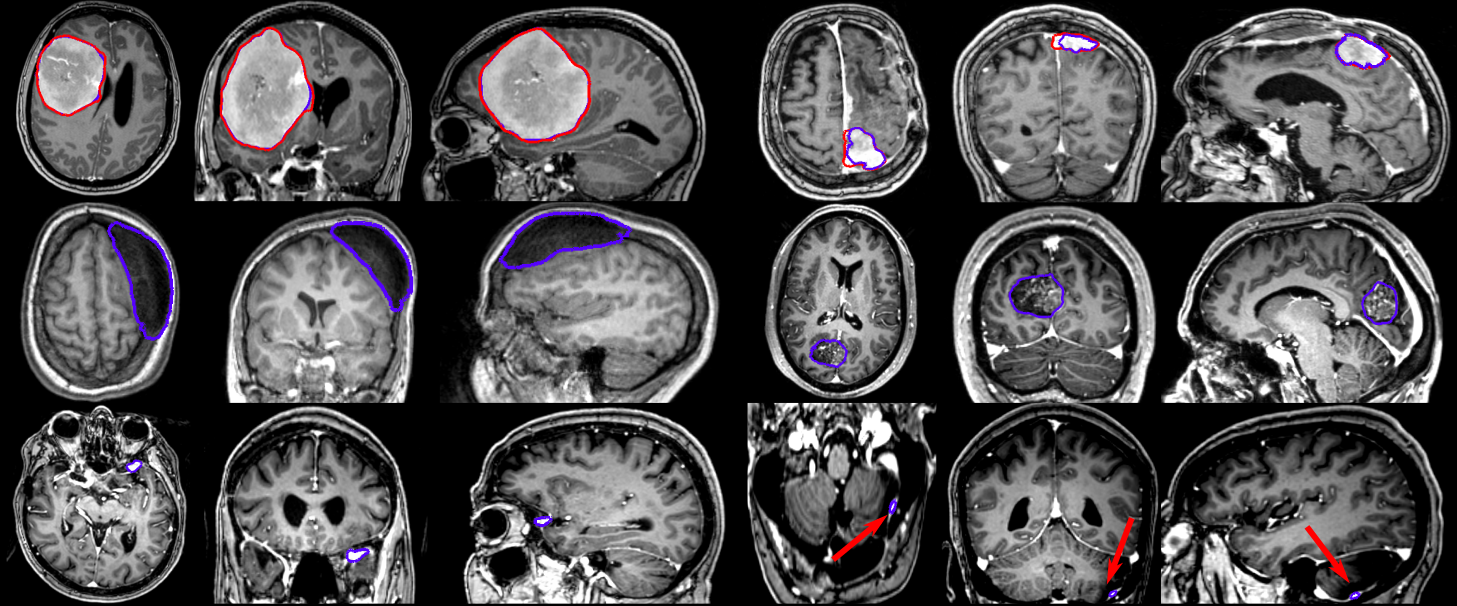}
\caption{Segmentation examples showing side-by-side the axial, coronal, and sagittal views respectively, where the automatic segmentation is shown in red and the manual annotation is shown in blue. The top row illustrates two properly segmented meningiomas with to the right a meningioma adjacent to the enhancing superior sagittal sinus and falx. The middle row shows to the left a non-enhancing meningioma exhibiting intraosseus growth and hyperostosis, and to the right a partly calcified, partly enhancing meningioma. The bottom row illustrates two meningiomas, with a volume smaller than 3\,ml, left undetected.}
\label{fig:seg-missed-cases-illu}
\end{figure}

Based on tumor volume, we grouped the meningiomas from our dataset into ten equally-populated bins and reported Dice performances for each bin using a box plot, as shown in Fig.~\ref{fig:results-volumeprov}. The average Dice score for the largest meningiomas, with a volume of at least 17.5\,ml, has not changed much across the four methods considered and hovers above 90\%. However, the number of undetected or poorly segmented large meningiomas is lessened with our best method, as can be seen by the reduced number of dots outside the whiskers of each box plot. With our best approach, we reach an overall recall of 93\%, which increases to 98\% considering only meningiomas larger than 3\,ml. We have identified eleven undetected cases with a volume larger than 3\,ml, and two examples are provided in the second row of Fig.~\ref{fig:seg-missed-cases-illu}. Both a non-enhancing intraosseus meningioma (to the left) and a partly calcified meningioma (to the right) are featured. All eleven cases are exhibiting some extent of contrast impediment compared to typical contrast-enhancing meningiomas (cf. first row of Fig.~\ref{fig:seg-missed-cases-illu}), which explains why our network struggles. Considering that the average meningioma in an hospital setting has a volume of $30.92$\,ml and the performances on meningiomas larger than 3\,ml, our proposed approach appears suitable and relevant.
The most significant results of our best appraoch can be observed for meningiomas smaller than 3\,ml, where the average Dice scores has been clearly improved. Starting from an average Dice of 46\% and recall of 62.7\% with PLS-Net, our best approach reaches an average Dice of 63.3\% and recall of 78.9\%. In Fig.~\ref{fig:seg-missed-cases-illu}, two representative meningiomas smaller than 3\,ml and left unsegmented by all methods, are illustrated in the third row. Locations around brain borders (e.g., eye socket, or brainstem) and close to larger blood vessels are especially challenging. In addition, using down-sampled MRI volumes with limited spatial resolution reduces such meningiomas to a very limited number of voxels the model can compute features from. Incidental findings of meningiomas' first appearance, when below 3\,ml, remains challenging and unreliable for broad clinical use. However, patients followed at the outpatient clinic have developed meningiomas of $7.62$\,ml on average, suggesting potential benefit from automatic segmentation using our models.

\FloatBarrier

\subsection{Speed performances study}
The model complexity, training convergence aspects, inference speed, and processing speed are reported in Table~\ref{tab:results-training-speed}. Multiple GPUs with slightly different specifications (e.g., memory clock speed and bandwidth) were used for training, and other CPU-demanding algorithms were episodically ran concurrently. As a result, the speeds per epoch and total train time reported cannot be directly and objectively compared but orders of magnitude are nonetheless relevant to consider.
The PLS-Net architecture is converging in less than 100 epochs, the fastest of all designs, but even with the smallest amount of total parameters its complex operations results in a total training time about three times longer than any full volume U-Net design for worse segmentation performances due to its shallowness (cf. experiments (ii) and (iii)). Training in a slab-wise fashion inherently increases the number of training samples which considerably lengthen the elapsed time per epoch by a tenfold compared to the fastest iterating design (cf. experiments (i) and (iii)). However, the convergence behaviour is not impacted as about 120 epochs are necessary, which is on-par with the various full volume designs such as experiment (vii). It is worth noting that while using accumulated gradients did not improve overall segmentation and detection performances, the models converge faster thanks to a better generalization from seeing more samples at every epoch (cf. experiments (vi) and (vii)). The combination of complex architectural designs and accumulated gradients enables convergence in about 110 epochs at best, which is equivalent to a more than reasonable total training time of 18 hours. Given the use of full volume inputs, the relatively small dataset size, and the quickly increasing total number of model parameters with advanced designs, one must trade carefully between model complexity and dataset size to prevent overfitting or similar convergence hurdles.\\
Regarding inference, doubling the number of parameters within a similar architecture does not alter the speed as can be seen between experiments (iv) and (ix), but only the shallow architecture from PLS-Net can go below the second. When running experiment (ix) on CPU, the inference speed reaches on average $8.66\pm0.09$\,seconds, slightly more than doubled compared to GPU usage. The largest gap between CPU and GPU usage happens when running experiment (ii).
With regards to the total processing time for a new patient's MRI volume, around 15\,seconds are necessary to provide segmentation predictions using a GPU, which would be fast enough not to hinder day-to-day clinical practice. Interestingly, and when considering computers deprived of high-end GPUs, the processing time on CPU remains similar with $15.39\pm0.15$\,seconds for experiment (ix). When running inference on the GPU for only one patient, the environment has to be initialized at first and the model loaded, making it speed-wise comparable with pure CPU usage. The serious bottlenecks when using computers with average specifications could be the RAM availability and the CPU parameters (e.g., frequency, number of cores).

\begin{table}[!t]
\caption{Model complexity, training convergence, and runtime performances for the different architecture designs studied, averaged across the five folds. All values in the table are reported with GPU support.}
\centering
\adjustbox{max width=\textwidth}{
\begin{tabular}{l||r|r|r|r||r|r|}
Experiment & \# params & s.$epoch^{-1}$ (s) & Best epoch & Train time (h) & Inference (s) & Processing (s)\tabularnewline
\hline
(i) UNet-Slabs~\cite{bouget2020fast} & 14.75\,M & $4\,103\pm313$ & $120\pm40$ & $160.2\pm44.3$ & $3.74\pm0.03$ & $15.53\pm0.16$ \tabularnewline
(ii) PLS-Net~\cite{bouget2020fast} & 0.25\,M & $1\,944\pm47$ & $91\pm23$ & $62.6\pm12.5$ & $0.92\pm0.01$ & $10.95\pm0.05$ \tabularnewline
(iii) UNet-FV & 5.89\,M & $374\pm4.7$ & $171\pm24$ & $21.0\pm02.5$ & $2.03\pm0.04$ & $12.12\pm0.08$ \tabularnewline
(iv) AGUNet-AG & 16.41\,M & $437\pm3.6$ & $138\pm28$ & $20.5\pm03.5$ & $3.88\pm0.04$ & $13.84\pm0.13$ \tabularnewline
(v) AGUNet-DS-AG & 16.41\,M & $434\pm3.4$ & $149\pm27$ & $21.7\pm03.3$ & $3.59\pm0.06$ & $14.13\pm0.21$ \tabularnewline
(vi) AGUNet-MS-DS & 18.66\,M & $472\pm3.5$ & $160\pm78$ & $25.0\pm10.3$ & $3.69\pm0.04$ & $14.35\pm0.15$ \tabularnewline
(vii) AGUNet-MS-DS-AG & 18.66\,M & $508\pm8.1$ & $120\pm29$ & $21.4\pm04.1$ & $3.71\pm0.04$ & $14.46\pm0.22$ \tabularnewline
(viii) DAUNet-MS-DS-AG & 25.72\,M & $434\pm3.3$ & $118\pm52$ & $18.0\pm06.2$ & $3.13\pm0.04$ & $13.85\pm0.17$ \tabularnewline
(ix) DAGUNet-MS-DS-AG & 30.96\,M & $476\pm3.3$ & $112\pm14$ & $18.9\pm01.9$ & $3.32\pm0.06$ & $16.13\pm0.28$ \tabularnewline
\end{tabular}
}
\label{tab:results-training-speed}
\end{table}

\section{Discussion}
\label{sec:discussion}
In this study, we investigated different deep learning architectures for segmenting meningiomas in T1-weighted MRI volumes, relying on attention mechanisms and global relationships.  
Directly leveraging an entire 3D volume has the clear benefit of boosting overall segmentation and detection performances, especially in terms of precision. Having access to global context and spatial relationships across the whole brain helps the model to better discriminate between contrast-enhanced meningiomas and bright anatomical structures (e.g., blood vessels), which drastically reduces the number of false positive predictions per patient. Nevertheless, the lack of satisfactory spatial resolution from the use of a down-sampled input volume can prove to be detrimental which rationalizes predictions using the slab-wise strategy to be more accurate pixel-wise on some occasions.
In order to build upon the strengths from each approach, while inhibiting their limitations regarding precision and pixel-wise segmentation accuracy, performing some extent of ensembling could bear potential. However, improved segmentation performance would come at the expense of the speed performance and additional complexity. The more models in the ensemble are, the longer training and inference computation takes.

\setlength{\parindent}{5ex} By extending a regular U-Net backbone architecture with various designs, we managed to further improve segmentation and detection performances. However, the only noticeable and clear contribution seems to come from the use of deep supervision. Setting trainable weights in the loss function to let the model learn how to best balance the loss from the probability map at each decoder level has not been attempted in this study. We hypothesize overweighting the coarse feature maps might favor recall while overweighting the fine feature maps would favor pixel-wise segmentation, and believe further investigation is of interest. Not supported by numbers, the effect of attention schemes has been qualitatively observed whereby predictions appear to be restricted to the brain itself or its close boundaries. From training examples, the model learned global spatial relationships to define some no-prediction zones where meningiomas are unlikely to occur. While such observations warrant a proper behaviour from the use of attention schemes, a greater variability in the training samples to feature meningiomas in all possible location might also be implied. Conversely, having witnessed some extent of brain-clipping effect using attention mechanisms can be considered as an indication for unsuitability towards meningioma segmentation. Given the possibility for meningiomas to potentially grow outward from every border of the brain, heavier preprocessing such as brain-masking used for glioma segmentation is inadvisable here as it would clip away parts of the tumor.
The use of multi-scale inputs also brought limited visible improvement, but the training samples fed to our architectures were already down-sampled from the original 3D MR volumes, starting the training with a degraded spatial resolution.

\setlength{\parindent}{5ex} For the time being, training our best architecture with the native MRI volume resolution is too challenging because of memory limitation on even high-end GPUs due to the sizable memory footprint. Nonetheless, working with down-sampled input volumes seems like the best trade-off solution as both recall and precision are favored. Detecting each and every meningioma accurately is critical as the actual pixel-wise segmentation task is more than often eased by the relatively good contrast and non-diffuse aspect of such tumors.
The current segmentation and detection performances are satisfactory for clinical use either as a tool for surgical planning or growth estimation. As indicated by an average Dice score above 90\% for meningiomas bigger than 3\,ml, when the average volume for patients having undergone surgery is 30.92\,ml, automatic measurements regarding the tumor aspect (i.e., volume, short-axis) and location (i.e., brain hemisphere and lobe) can be automatically generated.
For meningiomas with a volume below 3\,ml, somewhat worse performances were obtained. Detection of early meningiomas appears to be feasible but further improvements are needed for real and trustworthy use. As the average volume from patients followed at the outpatient clinic is 7.62\,ml, the current performances open for automatic and systematic growth computation and follow-up over time at a larger scale while reducing inter/intra-observer variability and being less time-consuming for clinicians.
The utmost challenging task remains the detection of tiny meningiomas exhibiting visual similarities with blood vessels, sometimes placed side-by-side or overlapping with them. The smallest meningiomas are also featured in a wider range of location (e.g., along the brainstem), and their total volume is only represented by a handful of voxels given the initial volume down-sampling. To address shortcomings from the latter, a finer down-sampling would help retain a superior spatial resolution but finding the proper balance between memory requirement and a prominent risk of overfitting would be challenging. Furthermore, broadening the dataset with additional samples featuring small meningiomas in a vaster range of locations might help the trained models generalize better. Alternatively, the use of other MR sequences such as FLAIR could help better distinguish between tumor and vessels. However, a larger panel of MR sequences might not be available at all time and processing only T1-weighted volumes makes our approach more generic and easy to use. Lastly, improving the architecture to make a better use of features available at the different input scales might be considered.

\setlength{\parindent}{5ex} To allow for exact comparison with the results from our previous study~\cite{bouget2020fast}, the dataset was not altered after the identification of outliers where meningiomas would not show with proper contrast, and which could be considered to be excluded from future studies. At it stands, the 11 outliers out of 600 volumes are additional noise during training and are a hindrance for the training process. By excluding them during validation, we would virtually reach 100\% recall with our best performing model for meningiomas bigger than 3\,ml.
In the validation studies, we chose to only rely on the threshold value PT, applied over the prediction map, to report the segmentation performance results. Given the almost perfect precision and high Dice scores obtained with the different full volume approaches, using an additional detection threshold, whereby a true positive is acknowledged only if the Dice score is above the given threshold, was not deemed necessary. Only few meningiomas have poor pixel segmentation and extent coverage (i.e., Dice score below 50\%), whilst the near-perfect precision ascertains the detection to be at least part of a meningioma.

\setlength{\parindent}{5ex} Even with sophisticated architectures and heavier designs, models are extremely fast to train and are converging in under 20 hours. Using an entire 3D volume as input compared to a slabbing strategy also speeds up training as less training samples are processed during each epoch. In addition, generalization schemes such as accumulated gradients help the model converge faster and reach a better optimum as can be seen by the reduction in standard deviation for the segmentation and detection measurements.
Interestingly, the inference speed is not heavily impacted by large variations in model complexity and these two parameters do not linearly correlate. Our dual attention guided architecture has 100 times more parameters than the shallow PLS-Net architecture, yet the inference speed is only multiplied by 3 reaching at most 3.7 seconds which is still fast enough and relevant for clinical use. The biggest hurdle for deployment in hospitals would be the large variability in hardware from low/mid-end computers and where shallower architectures like PLS-Net could thrive. The current disparity in performances, around 6\% F1-score difference, remains too high for such consideration at the moment and further investigation in that direction is warranted.

\section{Conclusion}
\label{sec:conclusion}
In this paper, we pushed forward the investigations around spatial relationships and global context for the task of meningioma segmentation in T1-weighted MRI volumes. Integrated into a regular U-Net backbone, we experimented with concepts such as attention mechanisms, multi-scale input, and deep supervision. Improved segmentation and detection performances have been demonstrated when moving from slab-wise to more sophisticated and complex approaches leveraging the entire 3D volume. Almost perfect detection results for clinically relevant meningiomas were obtained, whereas the smallest meningiomas, with a volume below 3\,ml, remained challenging given the limited spatial resolution and limited amount of voxels to compute features from.
In future work, special care should be brought towards the training dataset, as in many applications the bottleneck for improving performances lies in the data diversity more than the method's design~\cite{hofmanninger2020automatic}. Nevertheless, smarter handling of multi-scale features should be investigated, such as spatial pyramid pooling, to better leverage the raw spatial resolution. Alternative loss function designs, using adaptive weighting or new concepts, might also improve the pixel-wise segmentation, especially around tumor borders.

\subsection*{Conflict of interest statement}
The authors declare that the research was conducted in the absence of any commercial or financial relationships that could be construed as a potential conflict of interest.

\subsection*{Informed consent}
Informed consent was obtained from all individual participants included in the study.

\subsection* {Acknowledgments}
This work was funded by the Norwegian National Advisory Unit for Ultrasound and Image-Guided Therapy (usigt.org).

\bibliographystyle{unsrt}  
\bibliography{references}

\end{document}